1# Web Application for Collaborative Semantic Web Information Architecture

Massimiliano Dal Mas
*me @ maxdalmas.com**Abstract* — **In this paper is analyzed the prototyping of the information visualization on a Web Application for community purposes in a collaborative environment representing an evolution of the actual social networks like Facebook, Instagram, Twitter, Linkedin, VirgilioPeople, … The intent of this work is to identify the most common features of Web App. for the information visualization based on the Semantic Web and discuss how they support the user's requirements in a "collaborative" environment. A solution for the context-aware development of UI is based on "joint meaning" understood as a joint construal of the creator of the community contents and the user of the community contents thanks to the context and interface adaptation using the faced taxonomy with the Semantic Web. A proof-of concept prototype allows showing that the proposed methodological approach can also easily be applied to existing presentation components, built with different languages and/or component technologies.**

*Keywords* — **H.5.2.r User interface management systems, H.5.4.a Architectures, H.5.4.b Navigation, H.3.3.b Information filtering, H.3.3.d Metadata, I.2.4.k Semantic networks, I.2.12.b Internet reasoning services, H.5.3.c Computer-supported cooperative work, K.4.3.b Computer-supported collaborative work**I. INTRODUCTION

A Web Application (or Web App) is the fusion of the interactive and multimedia user interface functionality of desktop/mobile applications with traditional web-applications with data and multimedia contents, offering a high interactivity level. HTML5 is the current standard for delivering Web Applications, supported by all major browsers. [1, 2]

The use of Web App answers to the new functionalities that users demand to web applications with only one plug-in installation. Those kind of approach allow obtaining a more responsive feedback and a friendlier rich User Interface (UI), nevertheless it could complicate the client introducing in some cases a new software slice called client-engine between the user and the server.

A Web App respect to a conventional web application demands more tasks and duties to the client-side level, those for the information management and for the information visualization on the UI. An application becomes a Web App not for the technology with which it is built, but for its purpose, the most important concept is the "user centered design" and the approach to the prototyping.

Web App takes behavior from both, desktop/mobile and web applications. From the first one, it takes the standard menu terms (file, edit, save and delete, …), the tools menu and the direct manipulation of an object (like drag and drop). From the other one it takes the concept of design based on the content of the site, the information retrieval behavior and the possibility to open the user application having the user preferences everywhere.

All these paradigms have a very strong impact on the user. But we have to understand how we can use them. The first thing to do is analyzing the end user and his/her needs. At this stage the technology that will be use is not important. We have to analyze his/her working environment and the vocabulary used in his/her work.

There are some other things to remember. When we translate our analysis in a prototype, we have to remember that the user approaches a software first with his/her instinct and then with his/her rationality. This doesn't mean that we have to make something for dummies, but that we have to design an interface that is easy to approach. For example: designing Web App that are on the web being possible to found a paradigm to overcome the web standard to create a "desktop area" on the web so the user can immediately understand the different environment, thinking that this UI can help the user to have a correct approach with the application.

The information architecture has the most important role designing a Web App: we need to have a strong foundation based on a common ground. A Web App is an application built to make a specific task, for example to write document (Gdocs: http://docs.google.com/) or to design wireframe (Balsamiq: http://www.balsamiq.com/) and something like this. We have to think to a specific user, not a general user.

In this work is considered the community user both as creator of contents and as reader of those in a collaborative environment.

The paper is structured as follows. In Section 2 is clarified the theoretical background considered between Web App and the Communication Design. In Section 3 is shown the approach proposed for a faceted dynamic design. In Section 4

---

If you have read this paper and wish to be included in a mailing list (or other means of communication) that I maintain on the subject, then send email to: *me @ maxdalmas.com* is introduced the use of the concept of "joint meaning" on the topic. In Section 5 is described the architecture and the implementation of the developed prototype. Finally, Section 6 draws conclusion and suggest directions for future works.

## II. WEB APP AND COMMUNICATION DESIGN

### A. Information Visualization for the Semantic Web

The Information Visualization can support the collaborative environments by providing semantic features that make information more accessible enhancing their usability.

On the Web, the Information Visualization provides approaches for dealing with a large amount of summarized information with graphical interfaces techniques to manipulate search results. Few works deal explicitly with the problem of visualization in the Semantic Web, many of them refer to an ontological view.

### B. The Semantic Gap and the goal-direct approach

Users can use and interact with the services of the Semantic Web as a source of data. All kind of users, in the range between the pupils to the science researchers, can not solve the problem on how to interface with the Semantic Web through the training.

Basically, the user and the data models have different needs:
- data models are designed to maximize efficiency, orthogonality, and scalability;
- the user interfaces are designed to maximize usability, intuitiveness, discovery and personalization.

Between these two different ways of thinking you need a link that can be provided through Web App according to the HCI (Human Computer Interaction)

From the HCI we know that users want tools to customize and track their information space with the ability to "see" the information in different ways being able to contextualize the right tool at the right time. The Semantic Web can describe these needs while Web App cans instantiated appropriate interfaces to the user.

One important fact to remember is that if we have to represent complex information, we have to explain the meaning of the objects used in the application to the end user, either in a direct way or making the "meaning" of the object very simple to discover. And once decided what a certain interface object does, use it in a consistent way throughout all the screens of the application.

The taxonomies have an important role to help the developers, because they use a clear paradigm to manage data and term meaning. It is possible to apply them to a Web App and declare it to the users so that they know how to use the same "vocabulary".

### C. Web App for Information Visualization

For years data visualization had a secondary position in the graphical area. It was the kingdom of pie chart.

Data visualization has a lot of importance in the scientific field, where it is important to understand at the first glance the meaning of a lot of data without having to read through them: in this case visualization creates a map that represents numbers and words in way that makes aggregate information emerge.

That was developed for scientific data visualization with the information visualization theory with a lot of papers wrote on this topic.

Nowadays, with new tools and framework for developing UI like WPF (Windows Presentation Foundation) and HTML5 with CSS3 [1, 2], it is becoming easy to develop interesting data visualization also for "everyday" applications.

But this requires a big mind shift in the way it is designed the user experience; the designers as well the developers have to change the approach used to represent data and find more creative and meaningful way to represent complex data.

### D. Web App for Collaborative Semantic Web

The tools for collaborative Semantic Web should be designed to allow individual control of all kinds of information in a way that would be more significant. This by removing the arbitrary barriers created by applications that handle only certain types of information and record sets as predefined set of relationships, been possible to leave available to each user a view and a more efficient information interaction.

Web App can be a powerful way to merge the research objectives between the HCI and the Semantic Web domains on the information visualization overcoming the browser interaction limits.

Peter Morville, on his blog, https://intertwingled.org, pointed out how technology moves fast while evolution moves slow putting our attention to the social context of the users that need to "information retrieval" referring to Calvin Mooers. We have therefore needed of the methodology of Alan Cooper of the *Goal-Direct Approach* [3] that is based on observation and understanding of the needs and objectives of its size and relational archetypal: "Looking at things from the point of view of users is given a unique and impressive view, which opens new opportunities for creative design" [3]. As evolution of it Pillan and Sancassani [4] propose the *Style Aware Design* for "…asking us to take responsibility for the needs of all stakeholders of communication of the promoters, managers and users, finding ways to harmonize their needs of expression".

A lot of famous designers tell to us that the most important thing is designing simple interface. But simple is not the same of "for dummies". It's like something that follows our instinct. A lot of our action are not immediately rational, we move fast at the beginning of an action (like to retrieve an information from a system) and then we use our rationality to understand it.

The designer has to stay in the middle: the design must be easy to understand and must allow an easy start of operations, but then must allow the user to wonder through the design and use his intellect to understand the next steps.

It is very difficult to understand the borders and stay within them. There are some books and authors that mark this way. In the *Information Foraging Theory* Pirolli said that the way humans search information is the same as our ancient instinct to search food and adapt ourselves to the ambient to survive. Information in our daily life is essential to work and live. [5]

### E. Faceted classification Vs. Traditional classification

The Collaborative aims to share ideas, information, and create new knowledge.

The Semantic Web aims to share data between different applications, companies and communities.

Collaborative Web and Semantic Web can be a catalyst for the formation of communities of people with similar interests who exchange information and knowledge.

Many websites organize information using taxonomy with a hierarchy tree, while the semantic systems (such as RDF and OWL) are structures with a label directly on the chart. Each node of this graph represents a resource, while the arcs between nodes represent the properties of the resource. Move between nodes in a graph is different from moving in a hierarchical tree, since different people may use different paths to reach the same node.

While the standard web organizes its contents with a predefined structure, the co-operative systems on the web use a comprehensive approach with a graph structure in evolution.

In traditional classification systems (also called traditional *enumerative* taxonomies or systems), each item is classified under one and only one category. Giovanni M. Sacco defines taxonomies as traditional "encoders property" that, starting from the father-category create gradually daughters-categories by the addition of new properties [6], examples are the *system of Linnaeus* and the *Dewey decimal system*. It has a proper and *unique* place within a single framework, large and very deep hierarchy, and can be found through a stepped "*father category > son category*". This makes such classifications very rigid and conservative, because structurally closed, centralized and institutionalized these schemes do not allow it in the indexing phase, the inclusion by the classifier of a new category. Such a system is therefore *one-dimensional* (the method of cataloging is unique) and very *vertically* extended - but it is possible to search through its internal by different methods (however in a limited way, essentially: title, subject, and author).

In faceted systems (also called analytico-synthetic drafted by Ranganathan [7] - including graphs) the classes are not containers but physical descriptors (i.e. concepts and properties of concepts) and their relationship is primarily semantic. Under this feature, these systems leave the idea of a priori enumeration of all classes for a methodology that allows users to create the classes "on the fly" from some previously decided properties. They assign an item to multiple categories or parameters, each representing an aspect or side of the object (facetted). To the verticality of the traditional cataloging systems (i.e. excessive branching in depth of hierarchies), and their rigidity, the faceted classification offers a system of *horizontal* and *open classes* or *facets*, where each facet is descriptive of a property, or face of an object. For these peculiarities it was defined even as a *multidimensional classification*. [6]

This is why the analytic-synthetic systems should be extremely open and flexible, giving the user complete freedom to create from time to time the classes they need through the combination of concepts (blocks) and their primary properties or relations with other concepts. Pieces of information are blended in a hypermedia reality with increasing relations on different kinds of information: text, picture, video, audio, … In those way a stimulation of one "sensory modality", as looking at a picture, automatically triggers perception in a second modality, as reading a text, with the *Synaesthesia* from the Ancient Greek *syn* as set and *aisthēsis* as perception.

The idea is to make available to a metadata aggregation for each piece of information where the metadata (RDF) and ontologies (OWL) can localize the single local hosts to build on the fly the overall information according to the joint meaning between the user that create the piece of information and the user that read the overall information.

### III. FACETED DYNAMIC LAYOUT

### A. Faceted classification and Knowledge Management

Considering a faceted classification only as a theoretical apparatus coined by science books is limitative. This approach, in fact, is the formalization of a technique of communication that we often use in a wide range of contexts, from the organization of personal information to.

The faceted classification has important advantages over other systems in particular: *multidimensionality*, *persistence*, *flexibility* and *scale*. These features prevent the deterioration of repository avoiding that changes have negative repercussions on the information organization.

### B. Faceted organization of information and interface

Studies such as the Cooper [3] referred on the metaphor of Lakoff and Johnson are trying to move from the concept of "monocline grouping" (single nested level) suggesting that the analytical synthetic approach works better because it reflects the human mind. Specific studies lacks on the relationship between classification and human mind but some useful clues are provided. It is no coincidence that many types of software are moving today to analytic-synthetic approaches: the suites of Apple, Google Gmail, Windows, etc.
There are important repercussions on the concept of space at the interface. In a tree architecture is always possible to represent the path of the user and his current position in terms of class father to son, or category, from level *n* to level *n + 1* (so-called "bread crumbs" that are at the base also of very popular CMS applications such as the commercial Documentum, Vignette, RedDot, … as the open source CMS: WordPress, Joomla, Drupal, …).

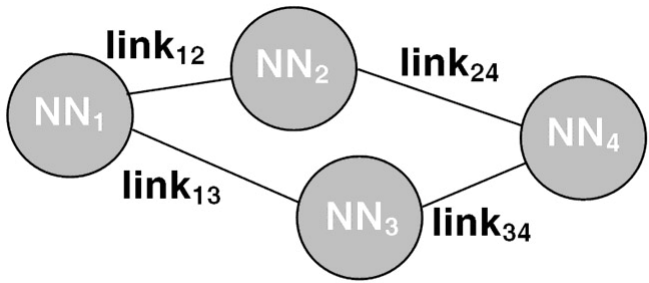

Figure 1. Nodes Navigation (NN) connected by links navigation used as different paths to achieve different goals by different users.

| SYSTEMS OF CLASSIFICATION | | |
|---|---|---|
| Organization System | HIERARCHICAL | FACETED |
| Navigation mode | Browsing | Filtering *and/or* Zoom |
| Space concept | Bread crumbs *or* space concept as position | Semantic relations among concepts |

Table I. Comparison of Hierarchical and Faceted systems of classification

A facet may well express itself in navigation but using research as the most efficient way to extract information.

Indeed the experiment with Web App for a collaborative environment does not deny the importance of navigation, but only highlights the priority of goals. Ways of navigation (WON) is the best way of navigating the path/s for users with certain profiles that need to complete their goal: this consists of "Nodes Navigation" (NN) connected by "links navigation" as a path where users don't care where they are but they care about where to go to achieve their current goal as shown in Figure 1.

In order for a single system to enable thousands of different users to achieve their hundreds of different goals, it needs a navigation system to support this unknown range of desires: on that Web App can help. Web App client side processing on data reduces the communication with the server.

According to the analytic-synthetic systems: the users who seen the contents can determine their own path through the aggregation / summary of various parameters.

That sense of space as a semantic relation between concepts may be the correct approach to the concept of space and movement within graphs: in these contexts the user can change from a two-dimensional to a multidimensional space changing hierarchies and relations. The position of the user is expressed by a relationship of concepts (or in RDF by subjects, predicates and objects): "I'm looking at a concept which has certain relations with other concepts".

The user will give value to what for him/her is significant, useful, and immediately understood, for those reasons the artwork should be semantic.

Table 1 compares the features of faceted and conventional hierarchical navigation, highlighting the potential of Web App.

## IV. JOINT MEANING

Communicating is a matter of performing certain types of actions. According to the Speech Act Theory [3, 4] what a speaker wants to communicate depends on his/her intention being a function of that.

According to Herbet Clark the meaning is jointly constructed by the *speaker* (user) and the *audience* (hearer) [8]. Community environment, like Facebook, are bringing the web on a more *Communication Acts* between a *speaker* and his/her *audience*.

The meaning of the communicative act produced by a users of the community, appears to be collectively constructed by the *speaker* (user that contributes to his/her own community space) and by his/her *audience* called by some communities as "friend" (that can see the speaker community space).

Communication involves not only the contributions of the *speaker* on the collaborative environment (with text, pictures, video, audio) but even how those are used by the *audience* according with the *faceted interface* used to visualize those.

Make something common between the speaker and the audience as having a common faceted visualization means communicate.

We can consider $U = \{1, ..., n\}$ be a finite set of *n users*, supposing that a user, $u$, communicates something to the other $n-1$ users of $U$. We can designate $u$ as the *speaker*, and the other $n-1$ users as the *audience*.

Joint meaning regarded as a joint construal of the *speaker u* and the *audience $U-\{u\}$*. For the proposed work joint meaning can be considered as joint activities of two or more subjects' users that can develop together a *faceted interface $\varphi$* according to their communication. [9 - 14]

Communication can be defined by as the fixpoint axiom of mutual belief that we can considered as the *fixpoint axiom of faceted interface for the communication*.

A Web App can be more that an "object" developed by one author and used by different kind of users, but it's an *aggregation experience* of people involved in developing and using this kind of communication as a Mapping Memory (http://www.alistapart.com/articles/mappingmemory/).

The experience, as the moment of the memory, is a composite dimension stressing on the cognitive, emotional and rational aspects being the encounter between what is going to happen - or the user can feel now - with what happened - or the user felt in the pass.

It is not possible to project an experience that a user can have, but is possible to set on the best way conditions for the user experience, so to "project for an experience".

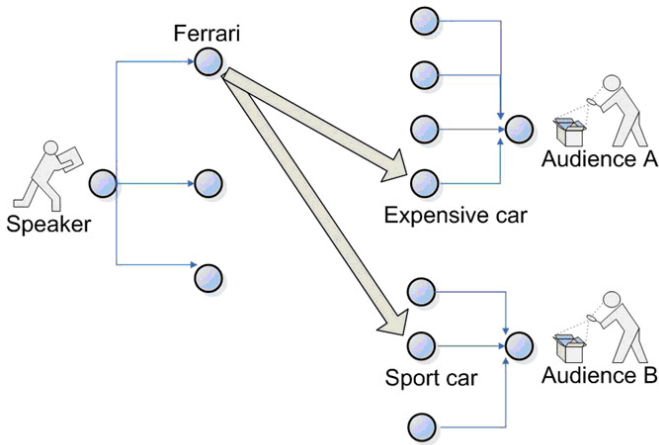

Figure 2. Faceted multiple matching: the system may aggregate some faceted elements into "Ferrari" tags for the *speaker* and "Sport car" for the *audience A* or "Expensive car" for *audience B* knowing their equivalence and can make reasoning on these tags.

Web App puts a new dimension for adaptive hypermedia system and personalized information visualization, instead of computing the adaptation steps at the server: Web App needs a client-side approach that can react immediately on user input.

Web App can allow the use of a "fluid" design where pages can be adapted to the tools and the customer choices thanks to the Semantic Web used to detect adaptation rules on Web App objects.

Rich Internet Clients (*RIC*) can directly execute all necessary adaptation steps based on a user model avoiding the latency of round-trips to the server by processing locally on the client.

## V. IMPLEMENTATION

Created for people community the prototype website has been developed locally with the main functionalities for testing the proposed approach. For the implementation was used the popular CMS WordPress with a plug-in that enables RDF and OWL output [10 - 17], and a themed AJAX interface is used to retrieve data integrating HTML5.

Faceted interface is defined in the Semantic Web by properties using triple to define the elements composing. Those are stored into the database access system, a SPARQL engines is integrated using Web App to combine the collaborative nature of Web2.0 with the ontologies of the Semantic Web.

Every faceted interface is composed by different kind of objects; those are identified by means of more than one aspect. So every object has more than one facet and the composition of every chosen facet compose the faceted interface between the *speaker* and the *audience* that can be identified by means of another facet. The faceted interface has a *joint meaning* in multiple domains as a reference designation of the interface with respect to the *speaker* and the *audience* being related to one facet, see Figure 2.

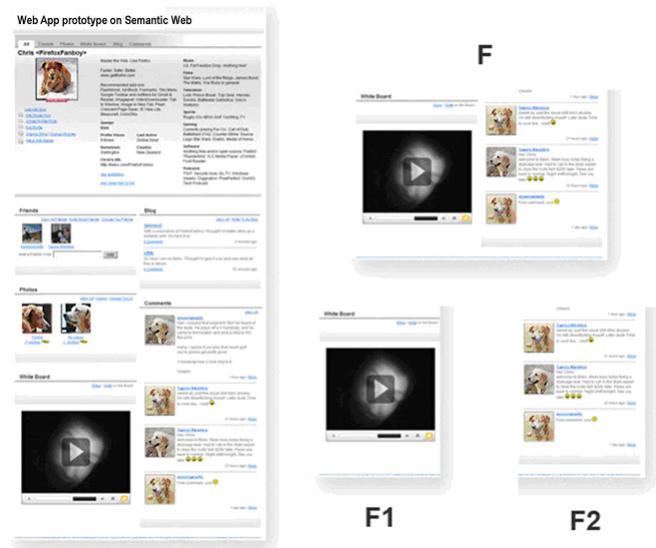

Figure 3. Prototype faceted interface, element F is a portlet seen as faceted element composed by 2 other portlets (F1 and F2) acording to the faceted sub-element chosen.

A centralized management of the identification register is used for the objects. Using the Semantic Web the metainformation referred to any object can be arbitrarily voluminous and structured, having any desired information granularity. Being flexible it is not required the use of long identification. So the identification can easily be kept stable over time; while at the same time the content of the metadata can be adapted to current needs (e.g. restructured, increase of granularity).

The information could be fragmented, put into NoSQL data bases (Elasticsearch and MongoDB), from which documents could be put together as needed including graphical presentations.

Instead of "smart" economizing with computing power it had become more essential to describe things logically and straightforward in order to enhance functionality, exchangeability and communication. Another very important requirement had become emphasized, namely that the reference designations should be possible to use over the entire life cycle of the "objects".

The faceted interface is constructed using algorithm based on the joint meaning. Ranganathan's theory [7] could help us to automatically determine intuitive facets that belong to either of intuitive and unintuitive categories; ontologies contain the knowledge on the kind of facet.

Acquisition of faceted subject metadata is done by a *Folksonomy social tagging* used as a means towards building such structure consolidated towards the Semantic Web. Folksonomy are related to every kind of contribution like text, picture, audio, video or free code that are considered as a *portlet* framework's UI. These portlets are made either manually, writing code by the user, or via a portlet creation wizard. Each portlet is a component or service with its own folksonomy for the faceted taxonomy and faceted interface, been customisable onto one or more faceted view aggregation in the page, see Figure 3, according to the *joint meaning*.

We use a set of pairs $(t_n, \varphi_m)$ to represent a faceted taxonomy: $t_1$ is a *tag* and $\varphi_1$ is a *faceted interface* [20].

$$(1) \quad F = \{(t_1, \varphi_1), \ldots (t_n, \varphi_m)\}$$

Each multiple association could be seen as a *superconcept* (2) defined using a *Semantic Enrichment Method*, consisting of a number of concepts that are equivalent with each other and a same tag matching between the Folksonomy and a domain ontology $O$ depending on the context of a user FOAF ontology (*Friend of a Friend*).

$$(2) \quad S \subseteq F \times O$$

To disambiguates multiple matching is used the *Superconcept Formation System (SFS)*, [19] a learning-based matching algorithm combined with rule based techniques that use a n-dimension Euclidean space vector for concepts with one semantic aspect; for each dimension is used an Artificial Neural Network technique to learn the weight and is calculated the weighted sum of dissimilarities from all corresponding dimensions.

## VI. RELATED WORK

The use of "joint meaning" to design collaborative UI is an underexplored area that arose from the pragmatics [9].
Different kinds of tools have been developed different aspects of the work. Two main classes are summarized below:

- Faceted Navigation

Most faceted navigation are used as interface for searching a large content database not considering different kinds of faceted visualization according to different kinds of audience (as done with "joint meaning") and using a fixed matching algorithm instead of a learning-based matching algorithm for automatic ranking of facet quality [20]

- Social Networks

Nowadays the Semantic Web is looked by Social Networks for providing their bones. Facebook is working to use the Semantic Web on social network data used to predict some individual private trait. [21].
Outside of Facebook, Microformats exist for tagging all kinds of information on ordinary web page. Microformats are simple standards that let you mark up elements in HTML documents to give them added significance, and to expose information to third party software and services.
Social Networks provides features for listing the people you know, publishing contact information, and advertising planning events for group of friends and not dynamically for the *speaker* and the *single audience user*.

## VII. EVALUATION

The evaluation was based on MiLE+ (Milano Lugano Evaluation Method) [22], which proposes an approach to usability evaluation under application-independent analysis (based on usability principles done by different experts) and application-dependent analysis (based on the requirements of the application, provide a step-by-step action guide for detecting the different problems with an assigned task, an example on Table II).

The overall results confirm that people prefer the faceted interface (83%), finding it useful (95%) and easy-to-use (82%).

EXAMPLE OF EVALUATION MATRIX

| Task: *Share a photo of a car between friends with same interest in cars* | PREDICTABILITY | UNDERSTANDABILITY | RICHNESS | COMPREHENSIBILITY | GLOBAL SCORE FOR THIS TASK |
|---|---|---|---|---|---|
| Scores | 8 | 8 | 5 | 6 | 6.75 (average score) |
| Weights | 0.1 | 0.1 | 0.5 | 0.3 | |
| Weighted Scores | 0.8 | 0.8 | 2.5 | 1.8 | 5.9 (weighted score) |

Table II. Only "*audience*" user with same interest in cars can see the picture of a car by the "*speaker*" user

## VIII. OVERVIEW AND CONCLUSION

With this paper has been shown how should be evolved present-day Information Architecture design methods to develop Web App application for the collaborative community environment. Web App with the Semantic Web can allow the use of a "fluid" design where pages can be adapted to the tools and the customer choice for the content and the layout.
Considering the relevant aspect of context-aware development of UI, it is shown how that requires increasing the expressive power of UI using Web App towards the "joint meaning" understood as a joint construal of the creator of the community contents (*speaker*) and the user of the community contents (*audience*) thanks to the context and interface adaptation using the faced taxonomy with the Semantic Web.

The described work has only scratched the surface of a huge problem being an initial step of a research program that will address several open issues:
1. a deeper comprehension of social commitment by subjects interaction as social reality intentionally constructed and how deontic affordances could be considered to produce joint meaning. [9] (see Carassa & Colombetti, 2009, for a first step in this direction);
2. enriching the methodological approaches by considering in depth different kind of Web App behaviours like: chat, multimedia synchronization, etc.;
3. developing automatic metrics for automatic facet ranking from the *Superconcept Formation System (SFS)*;
4. working on the semantics of the conceptual model, to enable automated methodological approaches;
5. distilling comprehensive guidelines supporting the design Web App;
6. measuring performance and optimizing the generated code;

77. continuing the industrial experimentation, by targeting other Web App platforms;
8. providing an UML profile and a visual notation for designing complex data-intensive Web application (WebML) for dealing with faceted interface using Web App for UI with the RUX-Method using the functionality of the existing web models. [23, 24]

ACKNOWLEDGMENT

I would like to thank Professor Marco Colombetti for his advice on Knowledge Engineering. I am especially indebted to all the reviewers' detailed comments and constructive suggestions on the manuscript.REFERENCES

[1] W3C Blog. World Wide Web Consortium (W3C) "HTML5 is a W3C recommendation", 2014 (https://www.w3.org/blog/news/archives/4167)
[2] W3C W3C Working Group Note 9 "HTML5 Differences from HTML4" 2014 (https://www.w3.org/TR/2014/NOTE-html5-diff-20141209/)
[3] A. Cooper, R. Reimann *"About Face 3 – The Essentials of Interaction Design"*. Wiley, 2007, p. 156
[4] M. Pillan, S. Sancassani "*Il bit e la tartaruga*" Apogeo 2004, p.122
[5] Pirolli, P. "An elementary social information foraging model", Proceedings of the 27th international conference on Human factors in computing systems: 605–614, 2009
[6] G. M. Sacco, Y. Tzitzikas "*Dynamic taxonomies aka Faceted Search*", Springer 2009
[7] A. Noruzi, "Application of Ranganathan's Laws to the Web". Webology, 2004, Article 8
[8] H. Herbert Clark *"Using Language"*. Cambridge University Press, Cambrdge 1996
[9] A. Carassa, M. Colombetti *"Joint meaning"*. Conditionally accepted for publication on the Journal of Pragmatics
[10] M. Dal Mas, "Folksodriven Structure Network". Ontology Matching Workshop (OM-2011) collocated with the 10th International Semantic Web Conference (ISWC-2011), CEUR WS vol. 814 (http://ceur-ws.org/Vol-814), 2011
[11] M. Dal Mas, "Elastic Adaptive Ontology Matching on Evolving Folksonomy Driven Environment" in Proceedings of IEEE Conference on Evolving and Adaptive Intelligent System (EAIS 2012), Madrid, Spain, 35-40, IEEE, (http://ieeexplore.ieee.org/xpl/articleDetails.jsp?arnumber=6232801) DOI: 10.1109/ EAIS.2012.6232801
[12] M. Dal Mas, "Elasticity on Ontology Matching of Folksodriven Structure Network". Accepted for the 4th Asian Conference on Intelligent Information and Database Systems (ACIIDS 2012) - Kaohsiung Taiwan R.O.C., 2012, CORR – Arxiv (http://arxiv.org/abs/1201.3900)
[13] M. Dal Mas, "Elastic Adaptive Dynamics Methodology on Ontology Matching on Evolving Folksonomy Driven Environment". Journal Evolving Systems 5(1): 33-48 (2014)
 (http://link.springer.com/article/10.1007%2Fs12530-013-9086-5)
[14] M. Dal Mas, "Modelling web user synergism for the similarity measurement on the ontology matching: reasoning on web user felling for uncertain evolving systems". Proceedings of the 3rd International Conference on Web Intelligence, Mining and Semantics, Article No. 17, ACM New York, NY, USA (http://dl.acm.org/citation.cfm?doid=2479787.2479799 ) ISBN: 978-1-4503-1850-1 DOI:10.1145/2479787.2479799
[15] L. Mussel (1964). Conformazione e struttura di bio-polimeri. Dip. di Scienze Chimiche – Univer-sità di Padova - Padova Digital University Archive.
[16] S. Corlosquet, R. Cyganiak, S. Decke, A. Polleres *"Semantic Web publishing with Drupal"*. DERI, Irelad, 2009
[17] *"IEC 81346-1 Ed.1: Industrial systems, installations and equipmen,t and industrial products – Structuring principles and reference designations."* IEC 2009
[18] S. Bindelli, C. Criscione, C. Curino, M. Drago, D. Eynard, G. Orsi *"Improving Search and Navigation by Combining Ontologies and Social Tags"* OTM Workshops 2008
[19] J. Huang, N. Huhns Michael, *"Superconcept Formation System–An Ontology Matching Algorithm for Service Discovery"* WWW Workshop 2006, Beijing, PRC
[20] J. Diederich, B. Wolf-Tilo, and U. Thaden *"Demonstrating the Semantic GrowBag: Automatically Creating Topic Facets for FacetedDBLP"* ACM IEEE Joint Conference on Digital Libraries 2007, Vancouver, Canada
[21] J. Lindamood, R. Heatherly, M. Kantarcioglu, and B. Thuraisingham. "Inferring Private Information Using Social Network Data", WWWC 2009
[22] D. Bolchini, F. Garzotto *"Quality of Web Usability Evaluation Methods: An Empirical Study on MiLE+"*; WISE 2007; Nancy (France)
[23] A. Bozzon, S. Comai, P. Fraternali, G. Toffetti: *"Conceptual modeling and code generation for rich internet applications"*. ICWE 2006: 353-360
[24] M. Linaje., C. Preciado., R. Morales-Chaparr, F. Sanchez-Figueroa *"On the Implementation of Multiplatform RIA User Interface Components"*. ICWE 2008: 44-49
**Massimiliano Dal Mas** is an engineer working on webservices and is interested in knowledge engineering. His interests include: user interfaces and visualization for information retrieval, automated Web interface evaluation and text analysis, empirical computational linguistics, text data mining, knowledge engineering and artificial intelligence. He received BA, MS degrees in Computer Science Engineering from the Politecnico di Milano, Italy. He won the thirteenth edition 2008 of the CEI Award for the "best degree thesis" with a dissertation on "Semantic technologies for industrial purposes" (Supervisor Prof. M. Colombetti). In 2012, he received the "best paper award" at the IEEE Computer Society Conference on Evolving and Adaptive Intelligent System (EAIS 2012) at Carlos III University of Madrid, Madrid, Spain. In 2013, he received the "best paper award" at the ACM Conference on Web Intelligence, Mining and Semantics (WIMS 2013) at Universidad Autónoma de Madrid, Madrid, Spain. His paper at 2013 W3C Workshop on Publishing using CSS3 & HTML5 has been appointed as "position paper", Paris, France.